\documentclass[epj]{svjour}
\usepackage{graphicx}
\usepackage{amsfonts}
\usepackage{amssymb}
\usepackage{amsmath}
\usepackage{infwarerr}
\usepackage{ltxcmds}
\usepackage{epstopdf}
\usepackage{amssymb}
\usepackage{bm}
\usepackage{amsmath}
\usepackage{color}
\usepackage[T1]{fontenc}
\usepackage[utf8]{inputenc}
\usepackage{widetext}
\usepackage{cuted}
\usepackage{ctable}
\usepackage{subfig}
\usepackage{graphics}
\begin{document}
\title{On the emergence of the ${\bf\Lambda}$CDM model from self-interacting Brans-Dicke theory in ${\bf d= 5}$}
\author{Luz Marina Reyes\inst{1}
\and Santiago Esteban Perez Bergliaffa\inst{2}
}                     
%
%
\institute{Departamento de Ciencias Computacionales, CUCEI, Universidad de Guadalajara. Av. Revoluci\'on 1500, 44430, Guadalajara 
 Jal., M\'exico, \email{luzmarinareyes@gmail.com}\and Departamento de F\'isica Te\'orica, Instituto de F\'isica, Universidade do Estado do Rio de Janeiro,\\
Rua S\~ao Francisco Xavier 524, Maracan\~a, Rio de Janeiro, Brasil, \email{sepbergliaffa@gmail.com}}
\date{Received: date / Revised version: date}
%
\abstract{
We investigate whether a
self-interacting  Brans-Dicke theory
in $d=5$ without matter and with a time-dependent metric
can describe, 
after 
dimensional reduction to $d=4$,
the FLRW model with accelerated expansion and non-relativistic matter.
By rewriting the 
 effective 4-dimensional theory as  
an autonomous three-dimensional dynamical system and studying 
its critical points, 
we show that 
the $\Lambda$CDM cosmology cannot emerge from such a model. This result suggests that 
a richer structure in $d=5$ may be needed
to obtain
the 
accelerated expansion as well as the matter content of the
4-dimensional universe. 
} 
\titlerunning{$\Lambda$CDM model from Brans-Dicke in $d=5$.}
\authorrunning{Reyes \& Perez Bergliaffa}
\maketitle
\section{Introduction}
\label{intro}
Several observations 
(such as SNe Ia, baryon acoustic
oscillations, and the cosmic microwave background, see for instance 
\cite{Alves2016}) indicate that the universe is currently 
undergoing an accelerated expansion. In the framework of the Standard Cosmological Model,
such an expansion is only possible if matter with unusual properties 
is added as a source of Einstein's Equations (EE) 
\cite{Frieman2008}.
The simplest candidate is the cosmological constant,
but there is a huge discrepancy between its theoretical value and the one that follows from observations \cite{Carroll2003}. Models with  scalar or
vector fields 
(see \cite{Li2011}  for a review of these and other candidates)
have also been considered to describe what is known as dark energy. Since none of these proposals is free of problems, 
several alternatives 
that 
 avoid the introduction of dark energy have been investigated. Among them we can mention theories of gravity that go beyond General Relativity \cite{Clifton2011} and inhomogeneous cosmological models \cite{Bolejko2016}. 
 Yet another interesting proposal is based on the hypothesis that the dimensionality of the universe is actually 
greater than four. The common 
theme in the many 
realizations of this idea is that
an effective energy-momentum tensor of purely geometrical origin,
generated by
the reduction of 
some theory of gravitation defined in $d>4$ to $d=4$,
is used to generate 
the accelerated expansion and/or ordinary matter.

In particular, the reduction of gravitational theories
from $d=5$ to $d=4$ has been repeatedly explored in the literature \cite{Wesson2006}. An appealing example of this type was presented in 
\cite{Wesson1992}, where the energy-momentum of ordinary matter in $d=4$ arises from 
the extra-dimensional sector of the theory defined by $G_{AB} = 0$. \footnote{Latin capital indices $A,B...$ go from 0 to 4, greek indices go from 0 to 3, and latin indices, from 1 to 3.}

More generally, theories in which 
the matter content in $d=4$ is induced by dimensional reduction of the  vacuum equations of a gravitational theory defined in $d=5$ 
are generically known today as Induced Matter Theories (IMT) \cite{Wesson2006}.
They 
have been extended in several directions, such as 
Brans-Dicke (BD) theory \cite{PoncedeLeon2009,Reyes2009,PoncedeLeon2010,Bahrehbakhsh2010,Rasouli2011,Rasouli2016}
\footnote{
For $d=5$ BD theory with matter see \cite{Qiang2004,Bahrehbakhsh2013}.},
$f(R)$ theories \cite{Borzou2009,Troisi2017}, and $f(R,T)$ theories \cite{Moraes2015}.  
Here we shall 
investigate the possibility of 
describing the accelerated expansion of the 4-dimensional universe as well as ordinary matter
starting from 
BD theory in the presence of a potential in $d=5$.
Cosmological evolution in self-interacting BD theory
has been studied both in $d=4$
(see for instance \cite{Santos1997,Sen2003,Chakra2009}) and in $d=5$ \cite{Periv2003}.
We shall show that in an appropriate cosmological setting, the $d=5$ self-interacting
BD theory  is equivalent to a 
self-interacting BD theory in $d=4$ plus an extra scalar field (associated to the time-dependence of the metric coefficient of the fifth dimension), which is suitable for the application of 
dynamical analysis methods. 
In particular, by imposing that the critical points of the dynamical system are deSitter-like, 
it is possible to determine whether the effective model
in $d=4$ can describe the accelerated expansion as well as the matter content of the 4-dimensional universe. 

The paper is organized as follows. In Sec. \ref{sec:level2}, we obtain
the effective theory in $d=4$ starting from a BD theory in vacuum in $d=5$ and in the presence of a potential.
In
Sec. \ref{sec:level3}, we write the field equations in $d=4$ as an autonomous three-dimensional dynamical system, and obtain its critical points,
under the assumption that $\dot H=0$. We pay special attention to the eigenvalues of the linearization matrix associated to each critical point,
and search for ranges of the parameters of the model such that the critical point is a stable one. 
We close with some 
comments in Sec. \ref{sec:level4}.

\section{\label{sec:level2}  Brans-Dicke Theory in ${\bf d=5}$ and its reduction to ${\bf d=4}$}

Our starting point is BD theory of gravity in five dimensions, with the action 
in the Jordan frame
given by
\begin{equation}
  \label{eq:FR1}
  ^{(5)}\!{\cal S}=\frac{1}{2\kappa_5} \int { d^{5}y \sqrt{^{(5)}\gamma}\left[ \phi\,^{(5)}\!R-\frac{\omega}{\phi} \gamma^{AB} \nabla_{\!A} \phi\nabla_{\!B}\phi-2V(\phi) \right]}
,
\end{equation}
\noindent where $^{(5)}\!\gamma$, is the determinant of the 5-dimensional metric $\gamma_{AB}$,
 $\phi$ is the BD scalar field directly coupled to the 5-dimensional Ricci scalar $^{(5)}\!R$,
 $\nabla_{\!A}$ is the covariant derivative in $d=5$,
 $\omega$ is the BD parameter and $V(\phi)$ is the scalar field potential.
The variation of the  action wrt  $\gamma_{AB}$ yields 
\begin{eqnarray}
  \label{eq:FR3}
  ^{(5)\:}\!G_{AB}&=\kappa_5\,^{(5)}T_{AB}+\frac{\omega}{\phi^2}\left[\nabla\!_A\,\phi\nabla\!_B\,\phi-\frac{\gamma_{AB}}{2}\nabla^C\phi\nabla\!_C\,\phi\right]+\nonumber\\
  &+\frac{1}{\phi}\left[\nabla_A\nabla_B\phi-\gamma_{AB}\,^{(5)}\Box\phi\right]- \frac{V(\phi)}{\phi}\gamma_{AB},
\end{eqnarray}
\noindent where 

$^{(5)}\Box=\nabla^{A}\nabla\!_{A}$, and 
 $^{(5)\:}\!G_{AB}$ is the Einstein tensor in $d=5$, given by $^{(5)\:}\!G_{AB}=\,^{(5)}\!R_{AB}-\frac{1}{2}\gamma_{AB}\,^{(5)}\!R$.

Variation of the action given in Eqn.(\ref{eq:FR1}) wrt $\phi$
 results in
 \begin{equation}
  \label{eq:BD3}
  \frac{2\omega}{\phi}\,^{(5)}\Box\phi- \frac{\omega}{\phi^2}\nabla^C\phi\nabla\!_C\,\phi+\,^{(5)}\!R- 2{V'(\phi)}=0,
\end{equation}
where the prime $(\prime)$ denotes derivative with respect to $\phi$.
 Taking the trace of Eqn.(\ref{eq:FR3}) we find
\begin{equation}
  \label{eq:FR4}
 ^{(5)}\!R=\frac{\omega}{\phi^2}\nabla^C\phi\nabla\!_C\,\phi+\frac{8}{3}\frac{^{(5)}\Box\phi}{\phi} +\frac{10}{3}\frac{V(\phi)}{\phi},
\end{equation}
which, when substituted in 
(\ref{eq:BD3}) yields
\begin{equation}
  \label{eq:BD4}
 ^{(5)}\Box\phi=-\frac{5V(\phi)}{3\omega +4}+\frac{3V'(\phi)}{3\omega +4}.
\end{equation}
We shall show next how 
Eqns.(\ref{eq:FR3}) and 
(\ref{eq:BD4}) 
are reduced to $d=4$ in a particular cosmological setting, giving as a result the usual BD theory with the addition of an
extra scalar field, whose dynamics and coupling to $\phi$ are determined by the reduction. \footnote{For a generalization of this procedure to an arbitrary number of dimensions see 
\cite{Rasouli2014}.}

In the coordinate chart $\left\{y^A\right\}=\left\{x^\mu,z\right\}$ we consider the 5D line element
\begin{equation}
  \label{eq:FR15}
  ds_{_5}^{2}=\gamma_{AB}dy^Ady^B=dt^2-a^2(t)(dr^2+r^2d\Omega^2)-\xi^2(t)dz^2,
\end{equation}
where $t$ is the time, $(r, \theta, \phi)$ are spherical coordinates on the hypersurfaces $t=$ constant, $z=$ constant, and $z$ is the coordinate along the
extra dimension, which we assume to be spacelike.
The metric
describing the standard cosmological model in $d=4$ is recovered
by restricting this line element to a hypersurface 
$\Sigma_0$
defined by $z=z_0
$=constant.\\
In order to obtain the effective field equations in $d=4$ from the dimensional reduction of Eqns.(\ref{eq:FR3}) and 
(\ref{eq:BD4}), the following 
expressions were employed:
%
\begin{subequations}\label{FR0}
\begin{align}
\nabla_{\mu}\nabla_{\nu}\phi=& {\cal D}_\mu{\cal D}_\nu \phi, 
\label{eq:FR17}\\
\nabla_{z}\nabla_{z}\phi=& -\xi\left({\cal D}_\alpha \xi \right)\left({\cal D}^\alpha \phi\right) 
,\label{eq:FR18}\\
^{(5)}\Box \phi  =&\Box \phi+\frac{\left({\cal D}_\alpha \xi\right)\left({\cal D}^\alpha \phi \right)}{\xi}
,\label{eq:FR19}\\
^{(5)}\!R_{\mu\nu}  =& R_{\mu\nu}-\frac{{\cal D}_\mu{\cal D}_\nu \xi}{\xi} 
,\label{eq:FR20}\\
^{(5)}\!R_{zz} = &  \xi\,\Box\xi
,\label{eq:FR21}
\end{align}
\end{subequations}
where
${\cal D}_\alpha$ denotes the 4D covariant derivative and $\Box={\cal D}^{\alpha}{\cal D}\!_{\alpha}$.
A long but straightforward calculation using all these expressions leads to the equations of the effective theory
in $d=4$ .
The equation for the BD field that follows from Eqn.(\ref{eq:BD4}) is
\begin{equation}
  \label{eq:BD5}
\ddot\phi+3H\dot\phi+\frac{\dot\xi}{\xi}\dot\phi=
-\frac{5V(\phi)}{3\omega +4}+\frac{3V'(\phi)}{3\omega +4}.
\end{equation}
From Eqn.(\ref{eq:FR3}),
with $A=B=0$, it follows that 
\begin{equation}
  \label{eq:BD94}
 3H^2+3H\frac{\dot{\xi}}{\xi}=\frac{\omega}{2}\left(\frac{\dot{\phi}}{\phi}\right)^2-3H\frac{\dot{\phi}}{\phi}
 -\frac{\dot{\xi}}{\xi}\frac{\dot{\phi}}{\phi}-\frac{V(\phi)}{\phi}.
\end{equation}
\noindent 
The spatial components of
Eqn.(\ref{eq:FR3}),
corresponding to $A = i$ and  $B = j$, can be written as
\begin{eqnarray}
\label{eq:FR83}
 2\dot{H}+3H^2+\frac{\ddot{\xi}}{\xi}+2H\frac{\dot{\xi}}{\xi}=\nonumber\\
= -\frac{\omega}{2}\left(\frac{\dot{\phi}}{\phi}\right)^2+H\frac{\dot{\phi}}{\phi}-\frac{V(\phi)(3\omega-1)+3\phi V'(\phi)}{\phi(3\omega+4)}.
\end{eqnarray}
Finally, setting  with $A=B=z$ 
in
Eqn.(\ref{eq:FR3}), we obtain
\begin{equation}
\label{eq:FR81}
  3\dot{H}+6H^2=-\frac{\omega}{2}\left(\frac{\dot\phi}{\phi}\right)^2+\frac{\dot{\phi}}{\phi}\frac{\dot{\xi}}{\xi}-\frac{V(\phi)(3\omega-1)+3\phi V'(\phi)}{\phi(3\omega+4)}.
\end{equation}
These equations reduce to
those presented in 
\cite{PoncedeLeon2010}, when the vacuum and homogeneous case is considered in the latter.
We shall show next that 
Eqns.(\ref{eq:BD5})-(\ref{eq:FR81}) can be written as an autonomous 3-dimensional dynamical system.

\section{Dynamical system}
\label{sec:level3}

In terms of the variables (see for instance \cite{Hrycyna2013hl})
\begin{subequations}
  \label{eq:FR36}
\begin{align}
x &= \dfrac{\dot\phi}{H\phi},\label{eq:FR37}\\
 y &= \dfrac 1 H \sqrt{\dfrac{V(\phi)}{3\phi}},\label{eq:FR38}\\
z &= \dfrac{\dot\xi}{H\xi},\label{eq:FR39}\\
\lambda &= -\phi \dfrac{V'(\phi)}{V(\phi)},
\label{eq:BD40}
\end{align}
  \end{subequations}
Eqn.(\ref{eq:BD94}) is written as
 \begin{equation}
  \label{eq:BD41}
y^2 = -1+\frac{1}{6}\omega x^2-\frac{1}{3}xz-z-x,
\end{equation}
and acts as a constraint. 
From Eqn.(\ref{eq:FR81}) 
it follows that
\begin{equation}
  \label{eq:BD42}
\frac{\dot H}{H^2} = 2x+2z-\frac{1}{2}\omega x^2+xz+\frac{3\,y^2}{3\omega+4}(\omega+\lambda+3).
\end{equation}
The actual dynamical system follows from Eqns.(\ref{eq:BD5})-(\ref{eq:FR83}), and it is given by
\begin{subequations}
  \label{eq:BD43}
\begin{align}
\frac{dx}{d\tau} =& -x\frac{\dot H}{H^2}-x^2-3x-xz-\frac{3(5+3\lambda)}{3\omega+4}y^2,\label{eq:BD44}\\
\frac{dz}{d\tau} =& -(z+2)\frac{\dot H}{H^2}-z^2+4x+z-\omega x^2+xz+ \nonumber \\ 
                 &+\frac{3(5+3\lambda)}{3\omega+4}y^2,\label{eq:BD45}\\
\frac{d\lambda}{d\tau} =& x\lambda\left[1-\lambda(\Gamma-1)\right],\label{eq:BD46}
\end{align}
  \end{subequations}
  where $\dfrac{d}{d\tau}=\dfrac{d}{d\ln a}$ and $\Gamma=\dfrac{V''(\phi)V(\phi)}{V'(\phi)^2}$ is assumed to be a function of $\lambda$.\\
  Table \ref{Tab1} shows the critical points of the system given by Eqns.(\ref{eq:BD42})-(\ref{eq:BD43}), under the assumption that $\dot H=0$, which corresponds to a deSitter expansion compatible with the latest observations, as mentioned in the Introduction. We shall discard the critical point $P_{1}$ since it leads to $y^2<0$. Points  $P_{3}$ and
  $P_{4}$ shall also be discarded because each of them is associated to a single value of $\omega$. Hence we shall focus the analysis on $P_{2\pm}$, $P_{5\pm}$, and $P_6$. 

\begin{table*}[htb]
\begin{center}
\begin{tabular}{c c c c c c }
\specialrule{.2em}{.1em}{.1em} 
Critical point&$~~~\lambda ~~~$&$~~~x~~~$&$~~~z~~~$&$~~~y^2~~~$& ~~~Restriction on $\omega$~~~\\ \hline 
 \noalign{\vskip 0.1in}
$P_{1}$&  $ -\frac{5}{3} $    & 0
   & 1 & -2   &  - \\ [1.5ex]
$P_{2\pm}$ &  $0$ &$\frac{-3\pm\sqrt{-15-12\omega}}{\omega+2}$   & $-x-3$  & $0$& 
$\omega\leq -\frac{15}{12}$  \\ [1.5ex]
$P_{3}$&  $0 $    & $-4$& $1$ & $0$  &  $\omega = -\frac 5 4 $ \\ [1.5ex]    $P_{4}$&   $0$     &$ -\frac{8}{3}$&$-\frac 1 3 $ & $0$ &
$\omega = -\frac{23}{16}$.
 \\ [1.5ex] 
$P_{5\pm}$&  $ \beta$ & $\frac{-3\pm\sqrt{-15-12\omega}}{\omega+2}$ & $\frac{ -6x(\omega+1)-9\omega-6}{ (3+x)(\omega+2)}$  &$0$ &  
$
\omega\leq -\frac{15}{12}$ \\ [1.5ex]       $P_{6}$&  $ -1$    & $\frac{ 1}{ \omega+1}$
       &$1$ & $-\frac 1 6\frac{12\omega^2+31\omega+20}{(\omega+1)^2}  $ & $-1.33\leq\omega\leq -1.25$  \\ [1.5ex]
\specialrule{.2em}{.1em}{.1em} 
\end{tabular}
\end{center}
\caption{{Critical points of the system given by Eqns.(\ref{eq:BD42})-(\ref{eq:BD43}) with $\dot H = 0$.
As explained in the text, only $P_{2\pm}, 
P_{5\pm}$, and $P_{6}$
will be considered in the subsequent analysis. The parameter $\beta$ is given by $\beta = \frac{1}{\Gamma -1}$.}}
\label{Tab1}
\end{table*}

We shall study next the
dynamical system given above by applying 
standard techniques, which include the introduction of new variables centered at the critical point, and the linearization of the system,
from
which 
it is possible to calculate the dependence of the Hubble parameter with powers of the expansion factor. Such powers will depend of the eigenvalues of the linearization matrix
at each critical point
(for details, see  \cite{Hrycyna2010} and references therein). 
Hence we shall begin with 
the analysis of the 
behaviour of the eigenvalues of the linearization matrix with $\omega$. The aim will be to 
obtain ranges for $\omega$ such that 
a given critical point is a stable node (for which all the eigenvalues must be real and negative), or a stable focus (characterized by one real and negative eigenvalue, and two  complex eigenvalues with negative real part). Hence,
only 
if the eigenvalues 
are such that their real part is negative for some range of values of $\omega$, we shall proceed with 
the calculation of $H(a)$.

The linearization matrix of the system in
Eqns. (\ref{eq:BD42})-(\ref{eq:BD43}) at a given critical point is given by
\begin{widetext}
\begin{equation}
A\, = \left[ \begin {array}{ccc}\left.A_{11}\right|_c&-{\frac {(\lambda_c-2\omega-1) {x_c}^{2}+2(3\lambda_c-3\omega+1) x_c+3(3\lambda_c+5)}{(3\,\omega+4)}}&-{\frac {\omega {x_c}^{3}+(z_c-3(\omega-2)){x_c}^{2}+6(z_c+2)x_c +9(z_c+1)}{2(3\,\omega+4)}}\\ \noalign{\medskip}\left.A_{21}\right|_c&-{\frac {(\lambda_c-2\omega-1)(\omega{ x_c}^{2}-2({ z_c}-1)x_c-12z_c)}{2(3\,\omega+4)}}&{-\frac{(1-z_c)\omega{x_c}^2+2({z_c}-1)(z_c+3)x_c+6({z_c}^2-1)}{2(3\,\omega+4)}}\\ \noalign{\medskip}{\lambda_c} \left( 1-{\lambda_c}\, \left( \Gamma \left( {\lambda_c} \right) -1 \right)  \right) &0&-\ 
\left.
\frac{d\Gamma(\lambda)}{d\lambda}\right|_{\lambda_c} {\lambda_c}^{2}{x_c}-2\Gamma( {\lambda_c} ) {\it \lambda_c}{x_c}+2{x_c}\,{\lambda_c}+{x_c}  
\end {array}
\right]
\label{matrix}
\end{equation}
\end{widetext}

where
\begin{small}
\begin{eqnarray*}
 \left.A_{11}\right|_c=& -\frac{1}{2(3\,\omega+4)}\left[ 4(3\omega-3\lambda_c-1)z_c+3({\lambda_c}-2{\omega}-1)\omega{x_c}^{2}+ \right. \\
        & \left. +(6(\omega-2)\lambda_c+34\omega+12+(-4{\lambda_c}+8\omega +4)z_c)x_c\right],
\end{eqnarray*}
\begin{eqnarray*}
\left.A_{21}\right|_c=-\frac{2}{3\omega+4}&\left[(\lambda_c-2\omega-1)(z_c-1)\omega x_c+\right] \\
        &\left.(-\lambda_c+\omega+1)({z_c}^2+2z_c-3).\right.
\end{eqnarray*}
\end{small}

We shall analyze next the behaviour of the eigenvalues
of this matrix at each critical point.

\subsection{${\bf P_{2\pm}}$}

Since $\lambda=0$ for these critical points,
it follows from the expression of the matrix $A$, given in Eqn.(\ref{matrix}), that
the eigenvalues do not depend of
the explicit expression of $\Gamma$. Hence, the results that follow will be valid for $\phi=\phi_c=0$ and $\left.(V'/V)\right|_{\phi_c}$ finite, or 
$V'(\phi_c)=0$.
\newpage
\subsubsection{$P_{2+}$}
\begin{figure}[h]
\subfloat{
{%
\setlength{\fboxsep}{0pt}%
\setlength{\fboxrule}{1pt}%
\fbox{
\includegraphics[width=8cm]{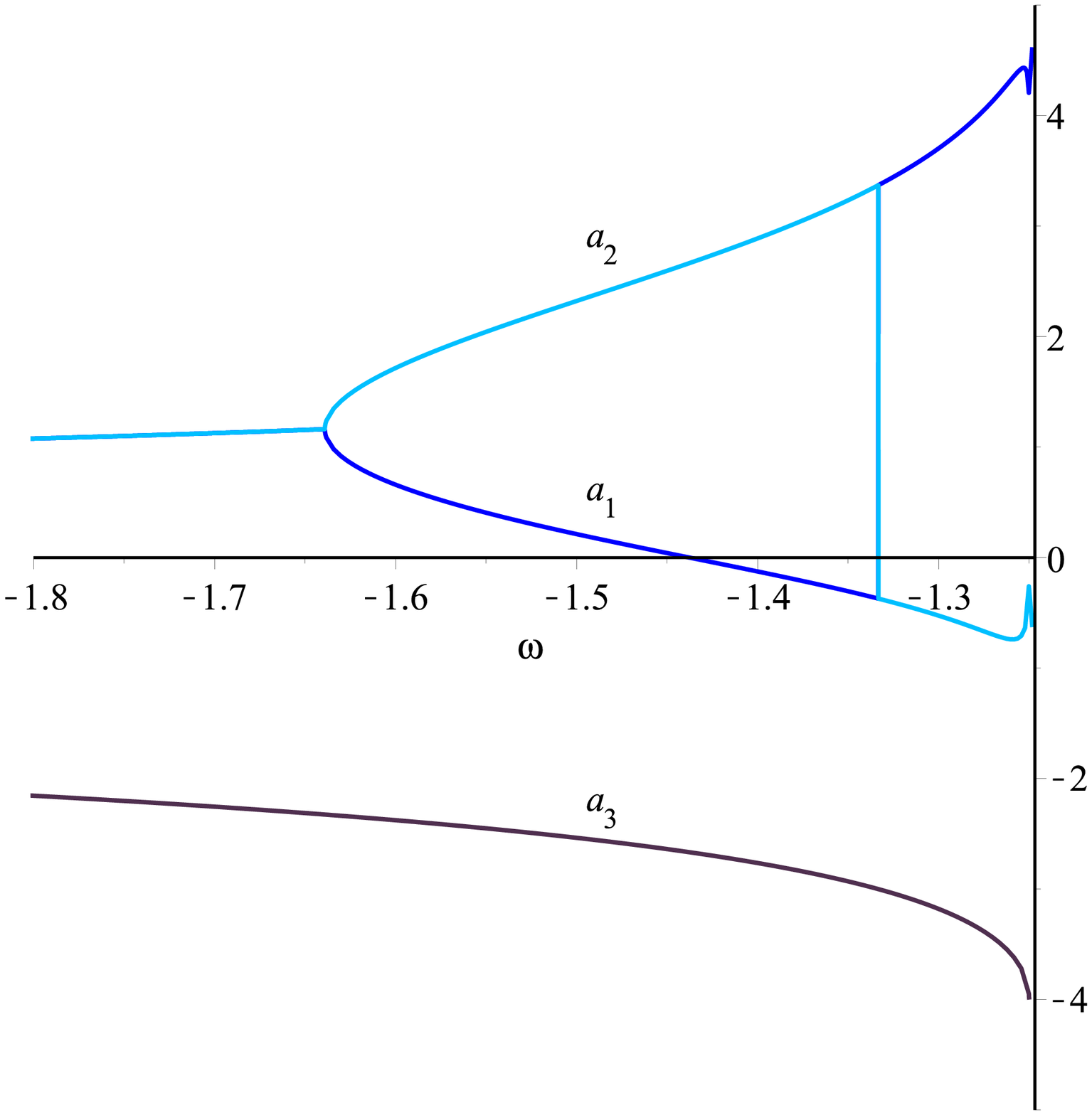}} \label{h1}}}
\caption{Real part of the eigenvalues corresponding to $P_{2+}$. 
The curves corresponding to $a_1$ and $a_2$ are superposed
to the left of approx. $\omega = -1.64$, and show a discontinuity at $\omega=-4/3$.
}
\label{rep2plus}
\end{figure}

The eigenvalues of the matrix $A$ for this critical point are given by
\begin{widetext}
\begin{eqnarray}
{ a}_1(\omega) & = & -\frac{1}{2(3\omega+4)(\omega+2)^2} \left[ -30\omega-24-9\omega^2+\sqrt{-15-12\omega} \: (10\omega+3\omega^2+8)-(-2304\omega^6+4116\omega^5+
\right. \nonumber\\
& & \left.
48906\omega^4+105600\omega^3+99600\omega^2+43776\omega+7296+\sqrt{-15-12\omega}\:(-2976\omega^5-10134\omega^4
 \right.\\ & & \left.
 -11424\omega^3-3888\omega^2
+768\omega+384) )^{1/2}
\right] ,
\nonumber
\end{eqnarray}
\end{widetext}
\begin{equation}
{ a}_2(\omega)=\frac{3-\sqrt{-15-12\omega}}{\omega+2}-{ a}_1(\omega),
\end{equation}
\begin{equation}
{ a}_3(\omega)=\frac{-3+\sqrt{-15-12\omega}}{\omega+2}.
\end{equation}
Fig. 
\ref{rep2plus}
shows the behaviour with $\omega$ of
the real part 
of each eigenvalue
associated to $ P_{2+}$. 
The plots show that there are no values of $\omega$ such that the real part of the three eigenvalues is real and negative. Consequently, $P_{2+}$ cannot be a stable point, and the behaviour of the system close to $P_{2+}$
cannot approach
the one currently displayed by the $\Lambda$CDM model. 

\subsubsection{$P_{2-}$}
\begin{figure}[h]
\subfloat{
{%
\setlength{\fboxsep}{0pt}%
\setlength{\fboxrule}{1pt}%
\fbox{
\includegraphics[width=8cm]{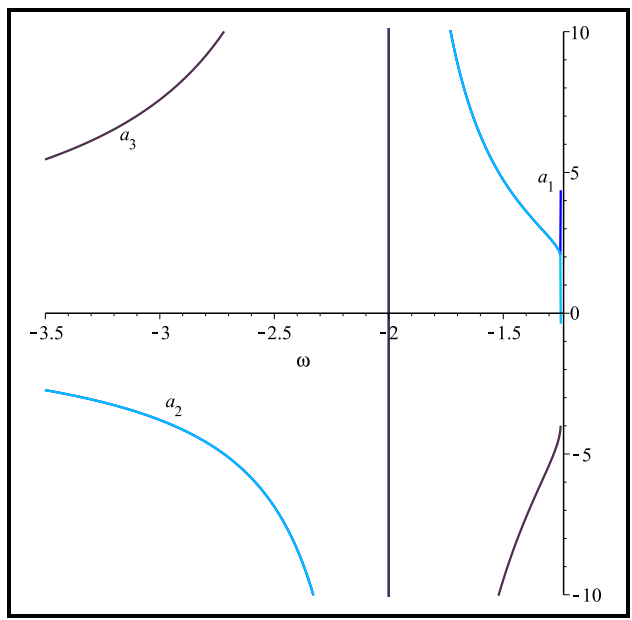}} \label{h1}}}
\caption{Real part of the eigenvalues corresponding to $P_{2-}$. The curves corresponding to $a_1$ and $a_2$ are not superposed
only near $\omega=-1.25$, and show a divergence for $\omega=-2$.
}
\label{rep2minus}
\end{figure}

The eigenvalues in this case are given by the following expressions:
\begin{widetext}
\begin{eqnarray}
{ a}_1(\omega) & = & \frac{1}{2(3\omega+4)(\omega+2)^2} \left[  \sqrt 6  (1216+7296\omega +16600\omega^2+17600\omega^3+8151\omega^4+686\omega^5 \right. \nonumber\\
& & \left.
-384\omega^6+\sqrt{-15-12\omega}\:(-64-128\omega+648\omega^2
+1904\omega^3+1689\omega^4
+496\omega^5) )^{1/2} \right.\\ & & \left.+9\omega^2+30\omega+24
+(3\omega^2+10\omega+8)\sqrt{-15-12\omega}\:
\right], \nonumber
\end{eqnarray}
\end{widetext}

\begin{equation}
{ a}_2(\omega)=\frac{3+\sqrt{-15-12\omega}}{\omega+2}-{ a}_1(\omega),
\end{equation}
\begin{equation}
{ a}_3(\omega)=-\frac{3+\sqrt{-15-12\omega}}{\omega+2}.
\end{equation}
Fig. 
\ref{rep2minus}
shows the 
plots of the real 
part of each eigenvalue associated to $ P_{2-}$. 
The plots show that there is no interval of values of $\omega$ such that the real part of the three eigenvalues is negative. 

\subsection{${\bf P_{5\pm}}$}

The critical points 
$P_{5\pm}$
depend of the potential through the condition $\lambda= \beta$. Since the eigenvalues for arbitrary values of $\omega$ and $\beta$ are given by long
algebraic expressions, we restrict here to the potential 
 $V(\phi ) = V_0 \phi^n$, such that 
 $\beta = \lambda = -n$ for every value of $V_0$ and $n$.
 This choice is justified by 
the fact that 
several effective quantum field theories can be related
to his kind of self–interacting potential \cite{Fujii2003}. 
 In particular, 
 we shall examine 
the cases $n=2$ and $n=4$, frequently considered in  cosmological scenarios (see for instance \cite{Sen2003,Boisseau2016,Carloni2008}).

\subsubsection{$P_{5+}$}

\begin{figure*}[h]
\subfloat{
{%
\setlength{\fboxsep}{0pt}%
\setlength{\fboxrule}{1pt}%
\fbox{
\includegraphics[width=7cm]{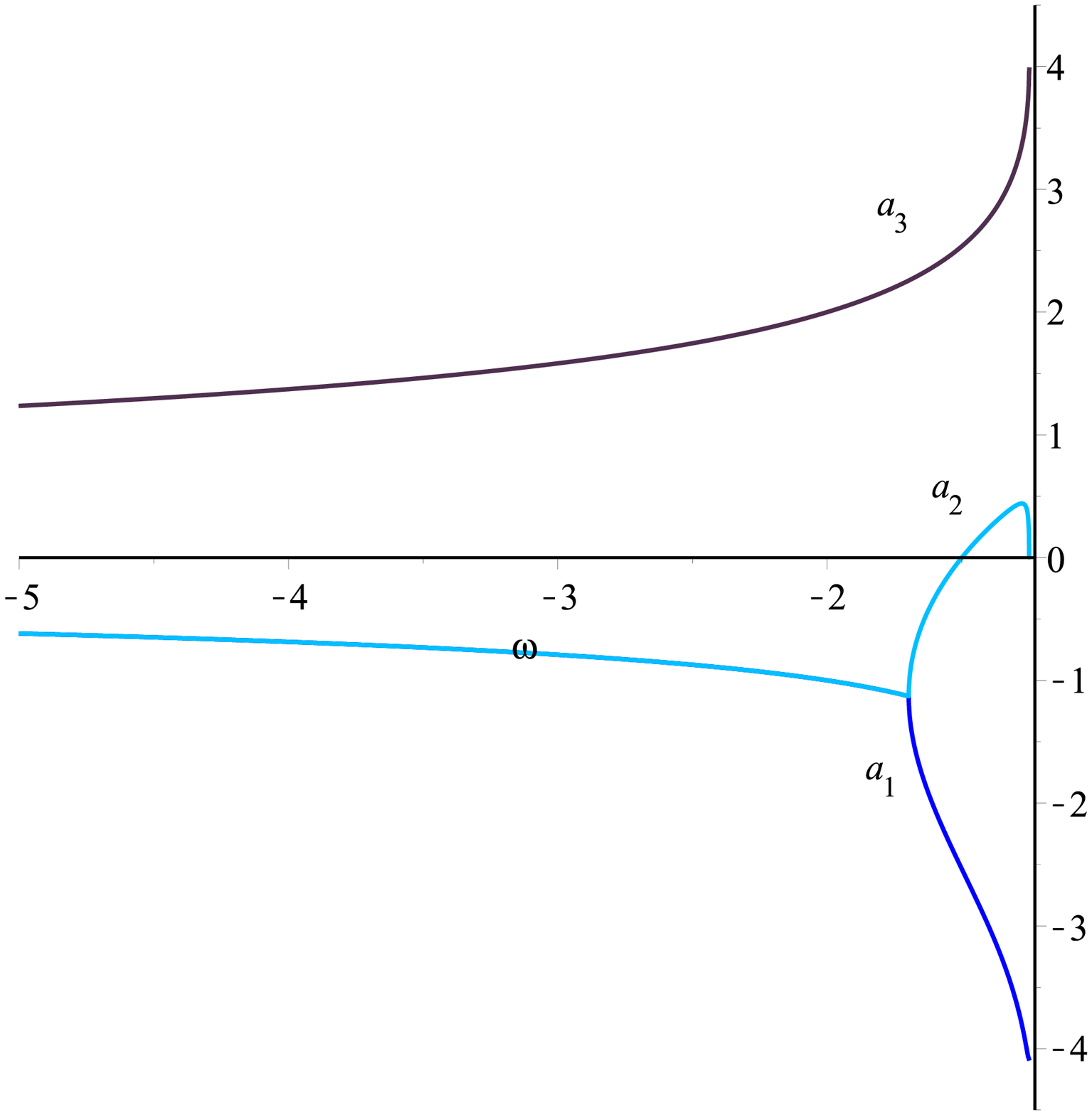}} \label{h1}}}
\subfloat{
{%
\setlength{\fboxsep}{0pt}%
\setlength{\fboxrule}{1pt}%
\fbox{
\includegraphics[width=7cm]{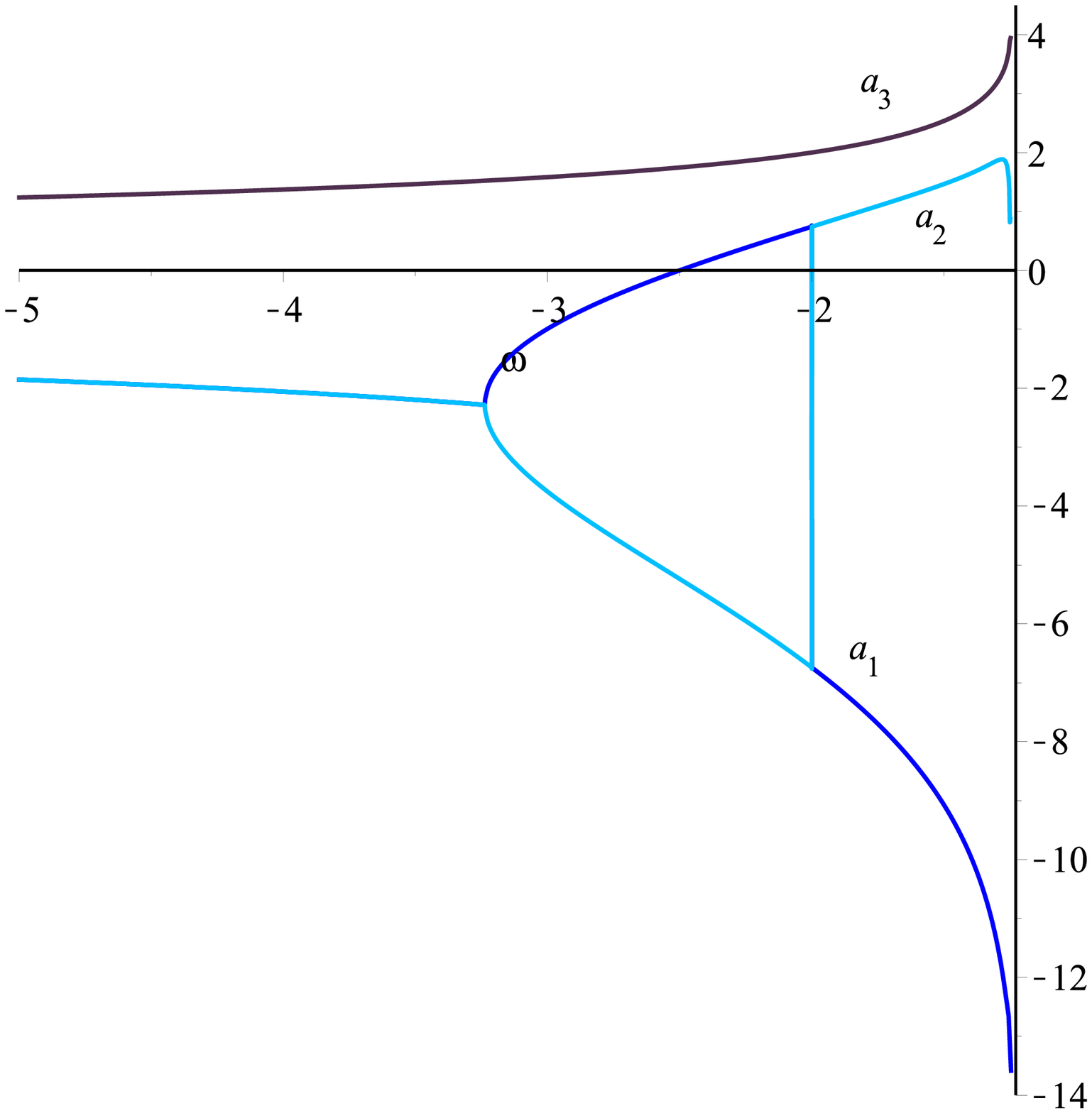}} \label{h2}}}
\caption{Real part of the eigenvalues corresponding to $P_{5+}$, for $n=2$ (left), and $n=4$ (right). The latter shows a discontinuity for $\omega=-2$.
The plots for $a_1$ and $a_2$ are superposed for $\omega\lessapprox -1.7$ (left) and $\omega\lessapprox -3.2$ (right)
}
\label{rep5plus}
\end{figure*}

The real part of the eigenvalues corresponding to the critical point $P_{5+}$
are plotted in Fig. \ref{rep5plus} for $n=2$ and $n=4$. None of the cases is associated to a stable critical point with $\dot H=0$.

\subsubsection{ $P_{5-}$}

\begin{figure*}[h]
\subfloat{
{%
\setlength{\fboxsep}{0pt}%
\setlength{\fboxrule}{1pt}%
\fbox{
\includegraphics[width=7cm]{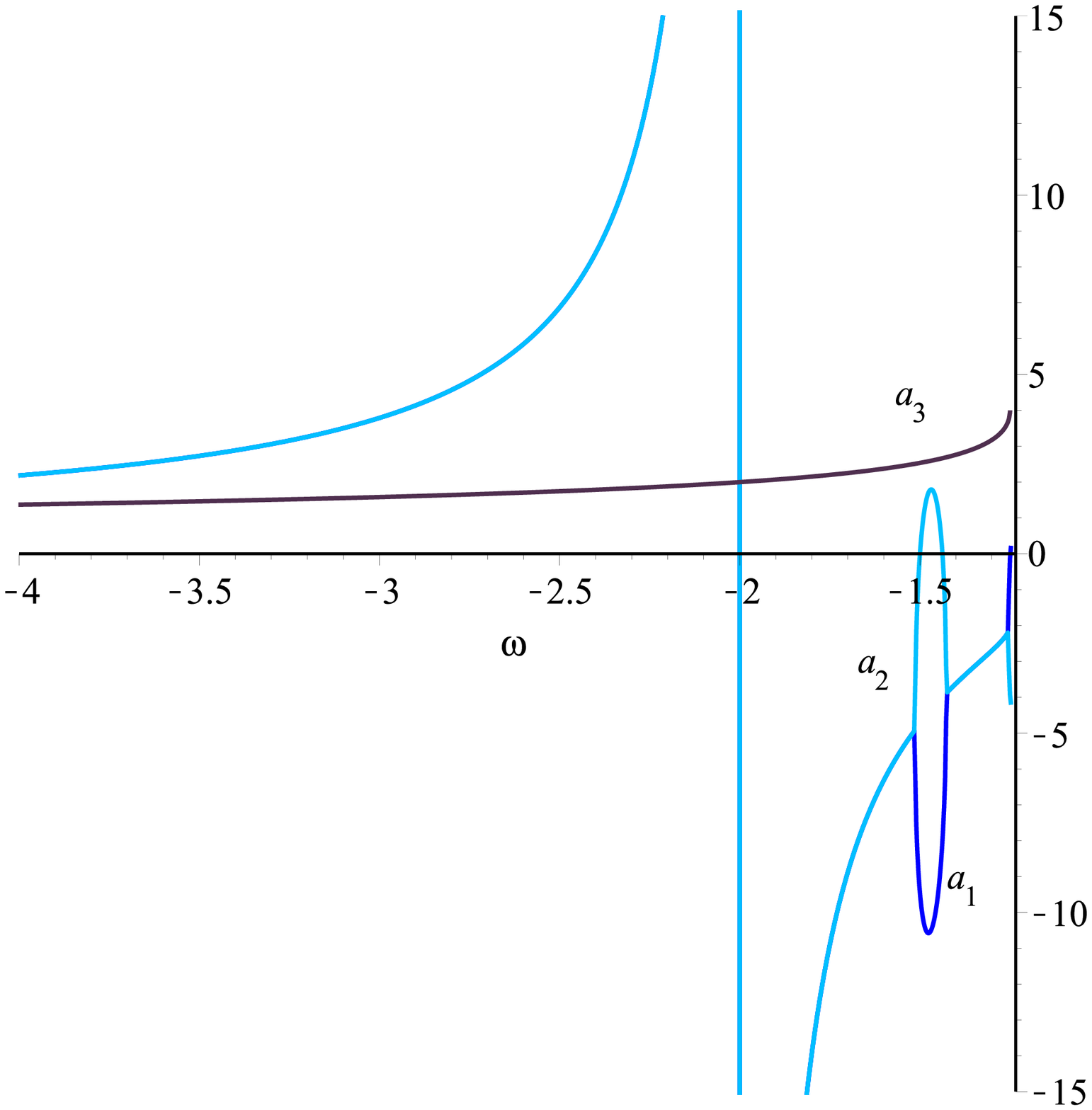}} \label{h1}}}
\subfloat{
{%
\setlength{\fboxsep}{0pt}%
\setlength{\fboxrule}{1pt}%
\fbox{
\includegraphics[width=7cm]{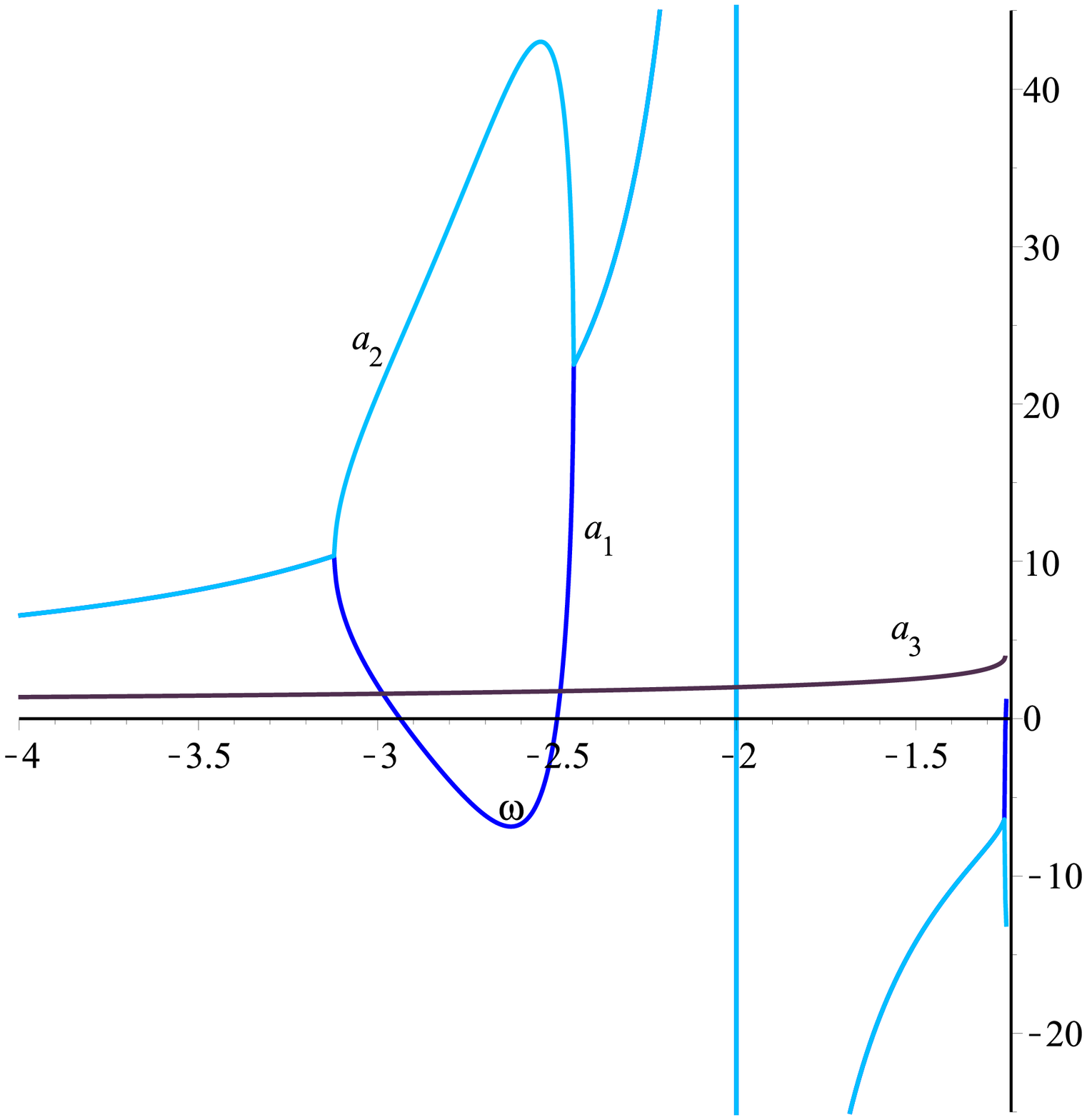}} \label{h2}}}
\caption{Real part of the eigenvalues corresponding to $P_{5-}$ and $n=2$ (left) and $n=4$ (right). $a_1$ and $a_2$ are singular at $\omega=-2$.}
\label{rep5minusn2}
\end{figure*}

The eigenvalues are plotted in Fig. \ref{rep5minusn2} for $n=2$ and  $n=4$,
and they fail to comply with the condition that their real part be negative. 

\subsection{ ${\bf P_6}$}
\begin{figure}[h]
\subfloat{
{
\setlength{\fboxsep}{0pt}
\setlength{\fboxrule}{1pt}
\fbox{
\includegraphics[width=8cm]{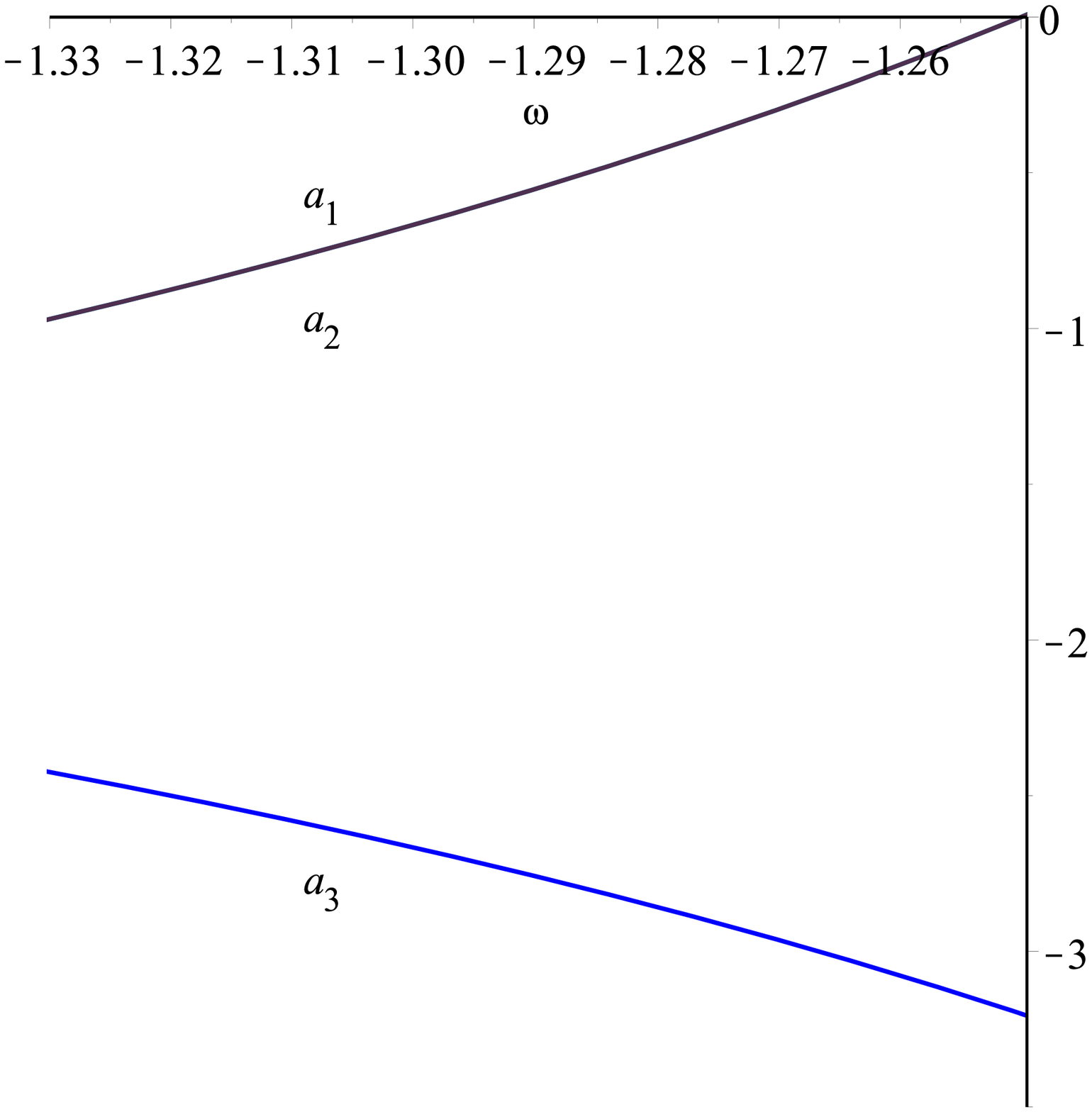}} \label{h2}}}
\caption{Plot of the real part of the eigenvalues corresponding to $P_{6}$, 
for
$\left.\frac{d\Gamma(\lambda)}{d\lambda}\right|_{\lambda_c}=-1.8$
.}
\label{repc6}
\end{figure}

The expression for the eigenvalues is in this case the following:
\begin{equation}
a_1(\omega)=-\frac{\left.\frac{d\Gamma(\lambda)}{d\lambda}\right|_{\lambda_c}+1}{\omega-1},
\end{equation}
\begin{equation}
a_2(\omega)=a_3(\omega)=-\frac{4\omega+5}{\omega-1},
\end{equation}
with $\Gamma (-1)=0$.
\footnote{Given any function $\Gamma(\lambda)$ such that $\Gamma(-1)=0$,
and $
\left.\frac{d\Gamma(\lambda)}{d\lambda}\right|_{\lambda=-1}=$ constant,
the explicit form of the potential can in principle be obtained 
from such a function and the definition of $\lambda$.
} The eigenvalues are shown in Fig. \ref{repc6} for
$\left.\frac{d\Gamma(\lambda)}{d\lambda}\right|_{\lambda_c}=-1.8$
. We see that, in spite of the fact that the real part of the three eigenvalues is negative, the eigenvalue $a_1$
could be associated to non-relativistic matter ({\emph {i.e.}}
is such that 
$Re(a_1)=-3$) only for a unique value of $\omega$.
Note that, although this conclusion follows from a particular value of 
$\left.\frac{d\Gamma(\lambda)}{d\lambda}\right|_{\lambda_c}$, the same will happen for any other value of the derivative
compatible with the restrictions, due to the specific form of the dependence of $a_1$ with the derivative. 
Hence, $P_6$ should also be discarded. 

\section{\label{sec:level4} Discussion}

We have examined whether a 4-dimensional universe in accelerated expansion and containing non-relativistic matter can be obtained by dimensional reduction of a self-interacting BD theory defined in $d=5$. The study
required rewriting the equations of the system as an autonomous 3-dimensional dynamical system.
The analysis of the eigenvalues of the linearized system shows 
that it has no 
stable equilibrium points subject to the condition $\dot H=0$, except for the critical point $P_6$, which 
is a stable critical point, but can 
describe 
non-relativistic matter only for a unique value of $\omega$ (given a value of
$\left.\frac{d\Gamma(\lambda)}{d\lambda}\right|_{\lambda_c}$
compatible with the restrictions)
. Hence,  the model cannot mimic the $\Lambda$CDM dynamics. This conclusion was obtained in full generality for  $P_{2\pm}$ and  $P_6$, and 
for $V(\phi)=V_0\phi^n$ 
and $n=2,4$
in the case of $P_{5\pm}$.
The failure of the model presented here
in describing both the accelerated expansion and the matter content of the 4-dimensional universe should perhaps be taken as an indication
that more complex models are needed, such as those in presented in
\cite{PoncedeLeon2010}, where the metric coefficient of the extra dimension is a function of both time and the extra coordinate.
We hope to 
go back to these ideas in a future publication.
\section*{Acknowledgments}
This work was supported 
by PROSNI 2015-2016,
PROFOCIE 2015-2016, P3E 235947 PROMOFID 2017, and Centro Universitario de Ciencias Exactas
e Ingenierias  of Universidad de Guadalajara.

%
 \bibliographystyle{unsrt}
%
\bibliography{bibliography}

\end{document}